\documentstyle[mprocl]{article}

\def\Journal#1#2#3#4{{#1} {\bf #2}, #3 (#4)}

\def\PRD{{\em Phys. Rev.} D}

\def\be{\begin{equation}}
\def\ee{\end{equation}}
\def\bea{\begin{eqnarray}}
\def\eea{\end{eqnarray}}
\begin{document}

%\title{THE ORIGIN OF STRUCTURES IN GENERALIZED GRAVITY}
\title{COSMOLOGICAL STRUCTURES IN GENERALIZED GRAVITY}

\author{J. HWANG}

\address{Department of Astronomy and Atmospheric Sciences \\
        Kyungpook National University, Taegu, Korea}


%%%%%%%%%%%%%%%%%%%%%%%%%%%%%%%%%%%%%%%%%%%%%%%%%%%%%%%%%%%%%%%%%
\maketitle\abstracts{
In a class of {\it generalized gravity theories} with general couplings
between the scalar field and the scalar curvature in the Lagrangian, 
we describe the {\it quantum generation} and the {\it classical evolution} 
processes of both the scalar and tensor structures in a simple and unified 
manner.}

%%%%%%%%%%%%%%%%%%%%%%%%%%%%%%%%%%%%%%%%%%%%%%%%%%%%%%%%%%%%%%%%%
\section{Generalized Gravity}

We consider a class of generalized gravity theories with an action
\bea
   S = \int d^4 x \sqrt{-g} \left[ {1 \over 2} f (\phi, R)
       - {1\over 2} \omega (\phi) \phi^{;a} \phi_{,a} - V(\phi)
       + \; L_m \right],
   \label{Action-GGT}
\eea
where $f$ is a general algebraic function of the scalar (dilaton) 
field $\phi$ and the scalar curvature $R$, and $\omega$ and $V$ are 
general functions of $\phi$; $L_m$ is an additional matter part of 
the Lagrangian.  
Eq. (\ref{Action-GGT}) includes the following generalized gravity 
theories as subsets:
(a) generally coupled scalar field,
(b) generalized scalar tensor theory,
(c) induced gravity,
(d) the low energy effective action of string theory,
(e) $f(R)$ gravity, etc.
Einstein gravity is a case with $f = R$ and $\omega = 1$.

%%%%%%%%%%%%%%%%%%%%%%%%%%%%%%%%%%%%%%%%%%%%%%%%%%%%%%%%%%%%%%%%%
\section{Cosmological Structures}

As a background world model we consider a spatially homogeneous and 
isotropic metric with $K = 0 = \Lambda$.
As the structures in this world model we consider the {\it most general} 
scalar, vector, and tensor perturbations 
\bea
   d s^2 
   &=& - \left( 1 + 2 \alpha \right) d t^2
       - a^2 \left( \beta_{,\alpha} + B_\alpha \right) d t d x^\alpha
   \nonumber \\
   & &
       + \; a^2 \Big[ \delta_{\alpha\beta} \left( 1 + 2 \varphi \right)
       + 2 \gamma_{,\alpha|\beta} + 2 C_{(\alpha|\beta)}
       + 2 C_{\alpha\beta} \Big] d x^\alpha d x^\beta.
   \label{metric}
\eea
The perturbed order variables $\alpha ({\bf x}, t)$, $\beta ({\bf x}, t)$, 
$\varphi ({\bf x}, t)$, and $\gamma ({\bf x}, t)$ describe the scalar type 
structure; the transverse vectors $B_\alpha ({\bf x}, t)$ and
$C_\alpha ({\bf x}, t)$, and the transverse-tracefree tensor
$C_{\alpha\beta} ({\bf x}, t)$ describe the vector and tensor structures, 
respectively.
Due to the high symmetry of the background model the three types of 
perturbations {\it decouple} from each other to the linear order
and evolve independently.
We will deal with only the {\it gauge invariant combination} of variables 
for the scalar and vector perturbations;
the tensor perturbation variables are naturally gauge invariant.

The {\it vector perturbation} is trivially described by a conservation
of the angular momentum:
For a vanishing anisotropic stress we have
\bea
   a^3 (\mu + p) \cdot a \cdot v_\omega \sim {\rm constant \; in \; time}, 
\eea
where $\mu (t)$, $p (t)$, and $v_\omega ({\bf x}, t)$ are the background 
energy density and the pressure, and the vorticity part of the fluid 
velocity in $L_m$.
The generalized nature of the gravity does not affect this 
result \cite{rotation}.
In the following we will concentrate on the scalar and tensor type 
structures.

%%%%%%%%%%%%%%%%%%%%%%%%%%%%%%%%%%%%%%%%%%%%%%%%%%%%%%%%%%%%%%%%%
\section{Classical Evolutions of the Scalar and Tensor Structures}

{}For the scalar field we let
$\phi ({\bf x}, t) = \phi (t) + \delta \phi ({\bf x}, t)$.
When we consider a scalar perturbation the following gauge invariant 
combination plays an important role \cite{GGT}
\bea
   \delta \phi_\varphi \equiv \delta \phi - {\dot \phi \over H} \varphi
       \equiv - {\dot \phi \over H} \varphi_{\delta \phi}.
\eea
$\delta \phi_\varphi$ is the same as $\delta \phi$ in the uniform-curvature
gauge ($\varphi \equiv 0$), and $\varphi_{\delta \phi}$ is the same as 
$\varphi$ in the uniform-field gauge ($\delta \phi \equiv 0$).
The perturbed action to the second order in the perturbation variables
can be arranged in a simple and unified form \cite{CT,GW}
\bea
   \delta^2 S = {1 \over 2} \int a^3 Q \left( \dot \Phi^2
       - {1 \over a^2} \Phi^{|\gamma} \Phi_{,\gamma} \right) dt d^3 x,
   \label{perturbed-action}
\eea
where for scalar and tensor perturbations, respectively, we have
($F \equiv \partial f/ \partial R$)
\bea
   \Phi = \varphi_{\delta \phi}, \quad
       Q = { \omega \dot \phi^2 + {3 \dot F^2 / 2 F}
       \over \left( H + {\dot F / 2 F} \right)^2 } \; ; \qquad
       \Phi = C^\alpha_\beta, \quad Q = F.
   \label{Phi-Q}
\eea
The non-Einstein nature of the theory is present in the parameter $Q (t)$.
The equation of motion becomes
\bea
   {1 \over a^3 Q} (a^3 Q \dot \Phi)^\cdot 
       - {1 \over a^2} \nabla^{2} \Phi = 0.
\eea
This has a general large scale solution 
\bea
   \Phi ({\bf x}, t) = C ({\bf x}) - D ({\bf x}) \int_0^t {dt \over a^3 Q},
   \label{LS-sol}
\eea
where $C ({\bf x})$ and $D ({\bf x})$ are the integration constants for
the growing and decaying modes, respectively.
This solution is valid for general $V(\phi)$, $\omega(\phi)$, 
and $f(\phi,R)$, and expresses the large scale evolution in a remarkably 
simple and unified form; results for the scalar structures are valid for 
the single-component subclass of gravity theories in (a)-(e) without $L_m$, 
whereas results for the tensor structures are valid for the general 
theory in eq. (\ref{Action-GGT}) including the additional matter 
contributions in $L_m$ (except for the transverse-tracefree stresses).

Notice that the growing mode of $\Phi$ (thus, $\varphi_{\delta \phi}$
and $C_{\alpha\beta}$) is conserved in the large scale limit
{\it independently} of the specific gravity theory under consideration.
Thus, the classical evolutions of very large scale perturbations
are characterized by {\it conserved quantities}.
[This conserved behavior also applies for sufficiently large scale
perturbations during the fluid eras based on Einstein gravity; in the 
fluid era the defining criteria for considering a perturbation to be 
large scale are the Jeans scale (sound horizon) for a scalar 
structure \cite{Fluid}, and the visual horizon for a gravitational wave.]
The integration constant $C({\bf x})$ {\it encodes the information about 
the spatial structure} of the growing mode.
Thus, in order to obtain the information on large scale structures, 
we need the information on $\Phi = C({\bf x})$ which must have been 
generated in some early evolutionary stage of the universe.

%%%%%%%%%%%%%%%%%%%%%%%%%%%%%%%%%%%%%%%%%%%%%%%%%%%%%%%%%%%%%%%%%
\section{Quantum Generations}

An acceleration phase in the early evolution stage of the universe
provides a mechanism which can magnify the ever existing microscopic
quantum fluctuations to macroscopic classical structures in the
spacetime metric.
In order to handle the quantum mechanical generations of the scalar
and tensor structures, we regard the perturbed parts
of the metric and matter variables as Hilbert space operators, $\hat \Phi$.
The correct normalization of the equal time commutation relation
follows from eq. (\ref{perturbed-action}) as
[in the quantization of the gravitational wave we need to take into 
account of the two polarization states properly, see \cite{GW}]
\bea
   [ \hat \Phi ({\bf x},t), \dot {\hat \Phi} ({\bf x}^\prime, t) ]
       = {i \over a^3 Q} \delta^3 ({\bf x} - {\bf x}^\prime).
   \label{commutation}
\eea
{}{\it For} $a \sqrt{Q} \propto \eta^q$ ($d\eta \equiv dt/a$) we have an 
exact solution for the mode function
\bea
   \Phi_{\bf k} (\eta) = {\sqrt{ \pi |\eta|} \over 2 a \sqrt{Q}}
       \Big[ c_1 ({\bf k}) H_\nu^{(1)} (k|\eta|)
       + c_2 ({\bf k}) H_\nu^{(2)} (k|\eta|) \Big], \quad
       \nu \equiv {1 \over 2} - q,
   \label{Phi-k-sol}
\eea
where according to eq. (\ref{commutation}) we have
$|c_2 ({\bf k})|^2 - |c_1 ({\bf k})|^2 = 1$; 
the freedom in $c_1$ and $c_2$ indicates the dependence on the vacuum state.
Although the condition used to get eq. (\ref{Phi-k-sol}) may look special,
it actually includes most of the proto-type inflation models investigated 
in the literature:
The exponential ($a \propto e^{Ht}$) and the power-law ($a \propto t^p$)
expansions realized in Einstein gravity with a minimally coupled 
scalar field lead to $\nu = {3 \over 2}$ and $\nu = {1 - 3p \over 2(1-p)}$, 
respectively.
The pole-like inflations ($a \propto |t_0 - t|^{-s}$) realized in
the generalized gravities in (b)-(d) with the vanishing potential
lead to $\nu = 0$; these include the pre-big bang scenario based on the 
low energy effective action of the string theory.

After their generations from the vacuum fluctuations during the acceleration 
era the relevant scale becomes the superhorizon size and the later classical 
evolution can be traced using the powerful conservation properties in 
eq. (\ref{LS-sol}); for the final observational spectrums of both structures 
generated in various inflation scenarios, see \cite{Infl,GW}. 

%%%%%%%%%%%%%%%%%%%%%%%%%%%%%%%%%%%%%%%%%%%%%%%%%%%%%%%%%%%%%%%%%
\section*{Acknowledgments}
This work was supported by the KOSEF, Grant No. 95-0702-04-01-3 and
through the SRC program of SNU-CTP.

%%%%%%%%%%%%%%%%%%%%%%%%%%%%%%%%%%%%%%%%%%%%%%%%%%%%%%%%%%%%%%%%%

\end{document}